\documentclass[conference,a4paper]{IEEEtran}

\IEEEoverridecommandlockouts

\usepackage{cite}
\usepackage{amsmath,amssymb,amsfonts}
\usepackage{algorithmic}
\usepackage{graphicx}
\usepackage{textcomp}
\usepackage{xcolor}
\usepackage{cleveref}
\usepackage{standalone}
\usepackage{subcaption}
\usepackage[inline]{enumitem}
\usepackage[super]{nth}

\usepackage{siunitx}
\usepackage{glossaries}

\setacronymstyle{long-short}
\newacronym{dnn}{DNN}{deep neural network}
\newacronym{stft}{STFT}{short time Fourier transform}
\newacronym{mae}{MAE}{mean absolute error}
\newacronym{mse}{MSE}{mean square error}
\newacronym{bce}{BCE}{binary cross entropy}
\newacronym{nmr}{NMR}{noise-to-mask ratio}
\newacronym{snr}{SNR}{signal-to-noise ratio}
\newacronym{dft}{DFT}{discrete Fourier transform}
\newacronym{ber}{BER}{bit error rate}
\newacronym{odg}{ODG}{objective difference grade}
\newacronym{peaq}{PEAQ}{perceptive evaluation of audio quality}

\def\BibTeX{{\rm B\kern-.05em{\sc i\kern-.025em b}\kern-.08em
		T\kern-.1667em\lower.7ex\hbox{E}\kern-.125emX}}

\begin{document}
	
	\title{Noise-to-mask Ratio Loss for Deep Neural Network based Audio Watermarking}	
	\author{\IEEEauthorblockN{Martin Moritz, Toni Olán, Tuomas Virtanen}
		\IEEEauthorblockA{
			\textit{Tampere University}\\
			Tampere, Finland \\
			firstname.lastname@tuni.fi}
	}	
	\maketitle
	
\begin{abstract}
	Digital audio watermarking consists in inserting a message into audio signals in a transparent way and can be used to allow automatic recognition of audio material and management of the copyrights. 
	We propose a perceptual loss function to be used in deep neural network based audio watermarking systems. The loss is based on the noise-to-mask ratio (NMR), which is a model of the psychoacoustic masking effect characteristic of the human ear. We use the NMR loss between marked and host signals to train the deep neural models and we evaluate the objective quality with PEAQ and the subjective quality with a MUSHRA test. Both objective and subjective tests show that models trained with NMR loss generate more transparent watermarks than models trained with the conventionally used MSE loss	
\end{abstract}

\section{introduction}

Digital audio watermarking consists in inserting a message into a host audio media, in a transparent way, meaning that the listener should perceive the degradation as little as possible and that their experience should not be affected. The retrieved hidden message can be used to prove the ownership of the media, or for identification and automatic management of the copyrights. The watermarks need to be robust against intentional or unintentional attacks such as compression, filtering or noise addition.  
The grounds for some traditional techniques like spread spectrum, patchwork, low-bit or phase codings, and echo-based watermarking were set~\cite{bender}  and have been developed further in the  following two decades as shown in the review~\cite{20_years_review} which  classifies about 70 solutions for audio watermarking. 

In the recent years, some \gls{dnn} audio watermarking solutions have emerged. 
The first~\cite{tegendal} is based on two \gls{dnn} networks: a U-net autoencoder to embed a binary message into the magnitude \gls{stft} of an audio segment and a deep convolutional multi label classifier for the extraction of the predicted message. The phase of the host \gls{stft} is directly reused for the watermarked \gls{stft}.

A  more recent solution~\cite{pavlovic} uses similar embedder and extractor networks for speech watermarking, with the important difference that input and output of the embedder are complex-valued  \gls{stft} and not only their magnitude, and both amplitude and phase of the watermarks are generated.

Those two previous works  use  the \gls{bce}  between the input and the extracted messages to guide the learning of the message retrieval, and the \gls{mae} to minimize the distortion between the input and the marked segments. The embedder and extraction networks are trained jointly and the two losses need to be weighted depending on the epoch: at the beginning the priority is given to the \gls{bce} to ensure the extraction of the message and afterwards the weight of the \gls{mae} is gradually increased to obtain a better transparency of the watermark. Another recent solution~\cite{liu2023dear}   uses an embedder, an extractor and a third discriminator network with an adversarial loss to improve the quality of watermarked audio, but it still uses the \gls{mse} loss to minimize the distortion.  In case of audio signal, loss functions like \gls{mse} or \gls{mae} minimize specific statistical criteria which do not match with the human capability to perceive signal modifications introduced by watermarks.

In this paper we propose to use a \gls{nmr}~\cite{brandenburg} based loss function for training \glspl{dnn}, which takes into account the psychoacoustic masking, to improve the perceptual transparency of the audio watermarks. \gls{nmr} is used as one of the statistics computed from audio signals from which the  \gls{peaq}~\cite{BS.1387} metrics has been calculated, and has been  applied for instance to minimize the distortion in  non-negative matrix factorization~\cite{nikunen}.

We do objective and subjective evaluations to show the better transparency of models trained with our novel \gls{nmr} loss in comparison to the \gls{mse}  commonly used to optimize the distortion between a \gls{dnn} generated  and an input signals.

We are solely interested in improving the transparency of the watermark thanks to the \gls{nmr} loss, and we are not addressing the robustness of the watermarking solution. Neither are we dealing with synchronization issue, and for the models we use the watermark needs to be retrieved exactly at the same position that it was embedded. To train a complete audio watermarking system, attacks can be performed on the embedded signals before the extraction. Those attacks can be \cite{pavlovic} random noise addition, low-pass filter, or sample suppression. To deal with the synchronization issue, traditional watermarking methods are commonly based on synchronization code \cite{shaoquan} which enables to retrieve efficiently the watermark location. In the case of \gls{dnn} solutions, synchronization attacks can be introduced to train the network to retrieve the message from another location than the one it was exactly embedded.  For instance \cite{Chen2023wavmark} finds a suitable location by brute force  and the usage of pattern bits.

The structure of the paper is as follows: Section \ref{section:method} explains the \gls{nmr} computation and gives the structure of the \gls{dnn} embedder and extractor for audio watermarking. Section \ref{section:evaluation} describes the audio datasets used for the training and the objective and subjective tests, explains how the models are trained and selected, how the tests are performed, and presents the results.

\section{proposed method}
\label{section:method}

The proposed loss function is applicable with any \gls{dnn} watermarking architecture but in this study we use an embedder and an extractor network  with similar architecture to~\cite{pavlovic}. 

\begin{figure}[t] 
	\begin{subcaptionblock}[T]{\columnwidth}
		\hfil \includegraphics[width=\columnwidth]{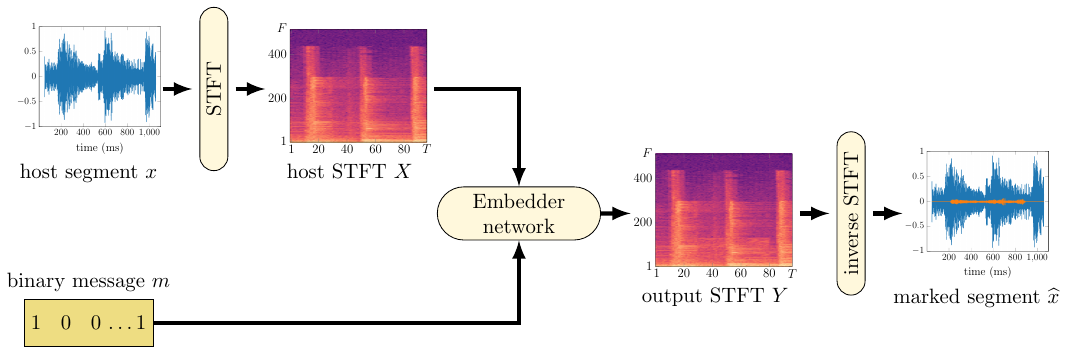} \hfil
		\caption{Embedder}
		\label{fig:insertion}
	\end{subcaptionblock}
	
	\begin{subcaptionblock}[T]{\columnwidth}	
		\hfil \includegraphics[width=\columnwidth]{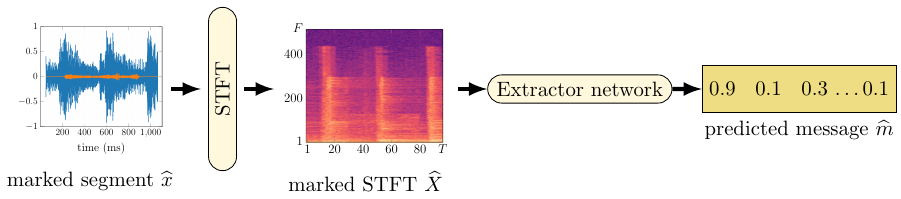} \hfil
		\caption{Extractor}
		\label{fig:extraction}
	\end{subcaptionblock}
	\caption{The two steps of the watermarking system}
	\label{fig:overall_structure}
\end{figure}

The embedder  (Fig \ref{fig:insertion}) takes as input a   host audio segment~$x$ and a 256-bit binary message~$m$. The \gls{stft}~$X$ of~$x$ is computed and fed with the message to the  \emph{embedder network}. 
The output of the embedder network is the marked \gls{stft}~$Y$ and the marked time-domain signal~$\widehat{x}$ is  obtained by its inverse \gls{stft}. 
The extractor (Fig \ref{fig:extraction}) takes as input the marked segment~$\widehat{x}$, computes its \gls{stft}~$\widehat{X}$ 
\footnote{The \gls{stft}~$Y$ output by the embedder network and the marked \gls{stft}~$\widehat{X}$ are not necessarily identical; even though the inverse \gls{stft} of~$Y$ is~$\widehat{x}$,~$Y$ might not be the \gls{stft} of any time-domain signal.}  
and feeds it to the \emph{extractor network} 
to get the the sequence of probabilities~$\widehat{m}$ which can be rounded to obtain the predicted message  of the same length, 256 bits, as the binary message~$m$.
The \gls{nmr} loss is computed between the marked and host \glspl{stft}~$\widehat{X}$ and~$X$.  

All the audio segments have a duration of \SI{1103}{\milli\second} and are sampled at~$F_S=\SI{44,1}{\kilo\hertz}$. 
The \gls{dft} length is~$N=1024$ samples, and we use the Hanning window and a time-overlap of 50\% to compute the \glspl{stft}. 
Therefore the \gls{stft} matrices have~$F=513$ rows and~${T=96}$ columns, corresponding   to the positive \gls{dft} frequencies    and the \gls{stft} time frames, and 2 channels for the real and imaginary parts. The 96-frame  duration  is chosen to match the decimation operations in the downsampling units of the extractor network.

\subsection{NMR loss}

The proposed loss used for the training of the \gls{dnn} is a combination of the noise-to-mask ratio (NMR) and the binary cross entropy (BCE) given as
\begin{equation}  \ell(\widehat{X},\widehat{m},X,m)  = \alpha  \text{NMR}(\widehat{X},X) + (1-\alpha)\text{BCE}(\widehat{m},m)\,,  \label{eqn:loss} \end{equation}
where~$\alpha$ is the transparency weight  between 0 (only the extraction matters) and 1 (only the transparency matters).

For computing the NMR in (\ref{eqn:loss}), we first need to compute the \emph{noise patterns}~$N_{c,t}$, i.e. the squared errors  between the marked \gls{stft}~$\widehat{X}$ and the host \gls{stft}~$X$, which are  decimated to the~$C=109$ critical band  using critical band mapping matrix~$U$, and weighted by outer and middle ear filter~$\omega$ as

\begin{equation}
	N_{c,t}  =   \sum_{f=1}^{F}	U_{c,f}     \left| \omega_f ( X_{f,t}-\widehat{X}_{f,t})  \right|^2  \;. \label{eq:noise_pattern} 
\end{equation}

Then the NMR is  the average of the ratios of the noise patterns and  the \emph{masking patterns} ~$M_{c,t}$ of the host  signal:
\begin{equation}
	\text{NMR}(\widehat X,X)  =   \frac{1}{C \cdot T} \sum_{c=1}^{C}  \sum_{t=1}^{T}  \frac{N_{c,t}}{M_{c,t}} \;.  \label{eq:NMR}
\end{equation}

Above,~$c$ is the critical band index,~$f$ is the \gls{stft} frequency index, and~$t$ is the time frame index.

\begin{figure}[t] 
	\hfil \includegraphics[width=\columnwidth]{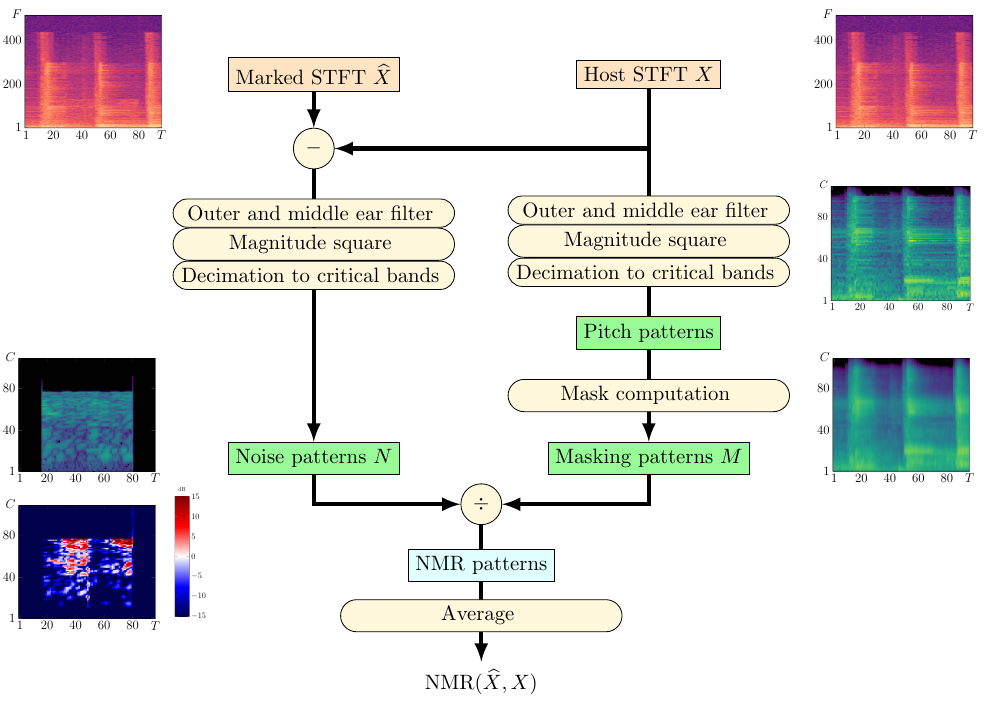} \hfil
	\caption{The computation steps of the \gls{nmr} }
	\label{fig:steps_nmr}
\end{figure} 

\begin{figure}
	\centering \includegraphics[width=0.9\columnwidth]{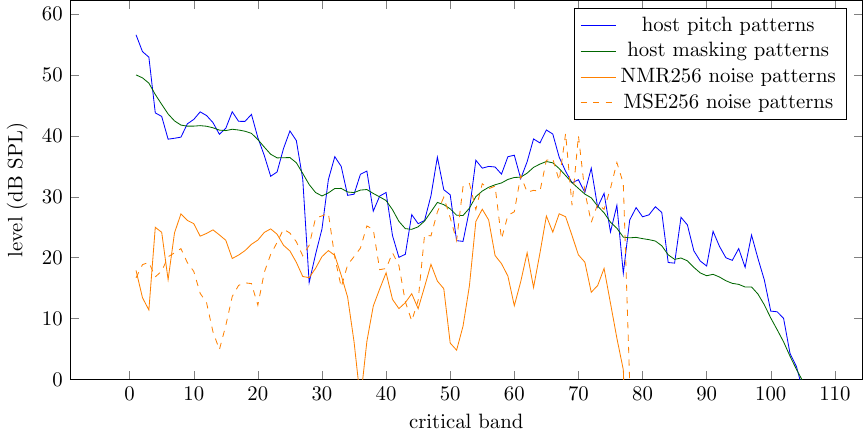}
	\caption{Pitch and masking patterns for a host signal, and the noise patterns for marked signals generated by two (NMR256 and MSE256) of our models}
	\label{fig:spectra}
	
\end{figure}

The detailed processing steps\cite{BS.1387} of \gls{nmr} calculation are illustrated on Fig.~\ref{fig:steps_nmr}  and summarized are as follows.
\begin{enumerate*}
	\item Both  complex-valued \gls{stft}~$X$ and~$\widehat{X}$ are computed and scaled by a loudness calibration constant as some factors of the model are level dependent.
	\item The error~$\widehat{X}-X$ is  weighted for each \gls{dft} frequency  by the frequency response~$\omega_f$  of the outer and middle ear filter and the  magnitudes are squared. 
	\item These squared errors are decimated according to the  109 critical bands to get  the noise patterns~$N_{c,t}$. The decimation is obtained by a  multiplication  by the \gls{dft}-to-critical bands mapping matrix~$U$.  The critical frequency bands range from \SI{80}{\hertz} to \SI{18000}{\hertz} with a constant resolution of~$0.25$  Bark. The rows of the matrix $U$ are the magnitude responses of band-pass filters equal to one inside the critical band, zero outside of it,  and  linearly interpolated in the transition bands. The columns of the matrix $U$ sum up to one.

	\item The host \gls{stft} is itself similarly scaled by the loudness calibration constant, weighted by~$\omega$ and the squared magnitude is decimated to the critical bands with the multiplication by the DFT-to-critical bands mapping matrix~$U$ to result  into the \emph{pitch patterns}. Those pitch patterns  are  spread first over the frequencies to model simultaneous masking and then in time to model temporal masking to get the excitation patterns, which are multiplied by a frequency-dependent offset to get the \emph{masking patterns}~$M_{c,t}$. The masking patterns correspond to the masking thresholds below which  a sound would be imperceptible. 
	\item The \gls{nmr} is the average of the ratios (\ref{eq:NMR}) of the noise patterns~$N_{c,t}$ and  the masking patterns~$M_{c,t}$.
\end{enumerate*}
The  pitch and the masking  patterns of a host signal in one example time frame, and the noise patterns between the host and marked signals for two models that we trained are shown on Fig.~\ref{fig:spectra}.

The \gls{nmr} calculation process is  detailed in~\cite{BS.1387} and some clarifications are given by~\cite{kabal} 
where  (\ref{eq:NMR}), converted into dB, is referred as the total NMR. In the ITU-R recommendations~\cite{BS.1387}, the error to compute the noise patterns is the squared difference of the magnitudes but here we compute the squared magnitude of the complex difference as shown in (\ref{eq:noise_pattern}).  
The reason for that lies in the fact that we want to use the \gls{nmr} as a loss to generate a marked signal and we need not only the amplitude but also the  phase to be as close as possible to the host signal. The sampling rate and STFT parameters we use are  different than in~\cite{BS.1387} where the signal is sampled at \SI{48}{\kilo\hertz}, the length of the DFT is 2048 samples.

\subsection{The Embedder} \label{section:embedder}

A~$128\times64$-submatrix of the host \gls{stft}~$X$, cropped between the \nth{4}  and \nth{131} frequencies and between the \nth{17} and the \nth{80} time frame, is  fed to the embedder network. 
The goal of the  frequency cropping is to limit the watermark in a frequency region (\SI{80}{\hertz}-\SI{5.6}{\kilo\hertz}) containing most of the audio content so that the watermark cannot be removed by simple filtering. The  cropping in time was intended to improve the extraction of a desynchronized watermark (which is not addressed in this study). As a consequence of it, there is no watermark at the beginning and at the end of the marked segments as shown~on~Fig.~\ref{fig:overall_structure}. 

\begin{figure} 
	\centering \includegraphics[scale=0.6]{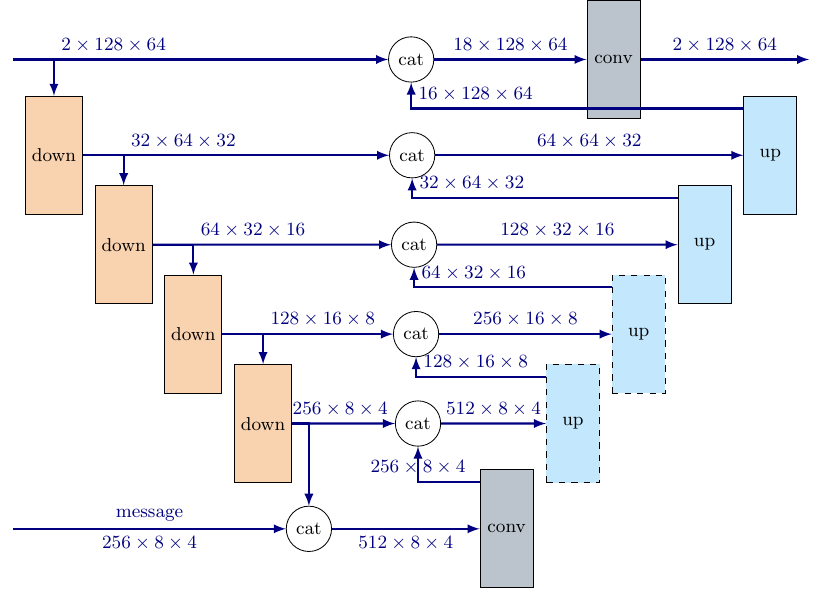} 
	\caption{The embedder network}
	\label{fig:embedder_network}
\end{figure}

As in~\cite{pavlovic} the embedder network (Fig.~\ref{fig:embedder_network}) is based on the U-net model and is composed of an encoder and a decoder. The encoder contains four downsampling units, each consisting of a convolutional layer, a 2D batch normalization layer and a leaky ReLU activation. Halving the width and the height of the feature maps is obtained with a stride of~$2\times2$ in the convolutional layers. The amount of  channels is increased from 2 in the input to 256 in the bottommost layer where the feature map has a size of 8$\times$4. This last output is concatenated along the channels dimension with the binary message where each of the 256 bit is replicated 8$\times$4 times to constitute  256 channels. This 512-channel concatenation is fed to an embedding unit composed of a convolutional layer of stride~$1\times1$ followed by a leaky ReLU activation, which outputs a feature map of 256 channels.

The decoder contains four upsampling units with skip connections coming from the encoder at each level. The input of the upsampling units are  the concatenation of the output of the corresponding downsampling unit and the output of the  unit below. The upsampling units are composed of a transposed convolution with a stride~$2\times2$, a 2D batch normalization layer and a ReLU activation. For the two bottommost ones, the activation is followed by a  dropout layer of probability 50\% as proposed in~\cite{pavlovic} to increase the robustness of the system against attacks.  The output of each upsampling unit has the same size (number of channels, height and width) than the input of the downsampling unit at the same level at the exception of the topmost unit which has 16 channels, whereas the network input feature map has two. Finally, those two feature maps are concatenated and fed to a convolutional layer. This upper skip connection is added in comparison to~\cite{pavlovic}, and another difference is that we have four downsampling and upsampling units whereas the embedder in~\cite{pavlovic} has one more of each and an intermediary feature map of 16 channels. 

The full output \gls{stft}~$Y$ is obtained by inserting the~${2\times128\times64}$ autoencoder output into the host \gls{stft} between the \nth{4} and \nth{131} frequency, and the \nth{17} and \nth{80} time frame. The inverse \gls{stft} of~$Y$ is returned by the embedder as the marked segment~$\widehat{x}$.

\subsection{The Extractor}

\begin{figure}
	\centering \includegraphics[scale=0.6]{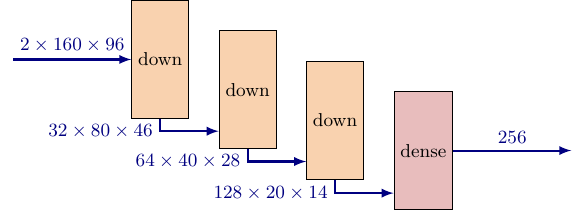} 
	\caption{The extractor network}
	\label{fig:extractor_network}
\end{figure}

The 160 lowest frequency bins and 96 time frames from the \gls{stft}~$\widehat{X}$ of the marked segment are fed to the \emph{extractor network} (Fig.~\ref{fig:extractor_network}). There is no cropping in time  to give the possibility to recover a desynchronized watermark.

The \emph{extractor network}  contains three downsampling units, each composed  of a convolutional layer with a stride of 2$\times$2, a 2D batch normalization layer, and a leaky ReLU activation. The final unit's output is fed to a dense layer and a sigmoid activation to compute the predicted message: a sequence of 256 floats, between 0 and 1. 

In both embedder and extractor, the convolutional and transposed convolutional layers have kernels of size  5$\times$5, and the ones followed by a batch normalization have no bias. The leaky ReLUs have a negative slope of \(0.2\).  

We also designed a second model to embed messages of length 512 bits. The only differences are the number of input channels of the convolutional layer in the embedding unit (768 instead of 512) and  the dimension for the output of the dense layer in the extractor (512 instead of 256).

\section{evaluation}
\label{section:evaluation}

We evaluate two models with message lengths 256 and 512 bits. 
Each is trained in two versions, one with the loss (\ref{eqn:loss}) that uses the \gls{nmr}, and the second one where \gls{mse} is used instead of the \gls{nmr} loss.  

\subsection{The audio data}

The models are trained with a subset of the FMA audio dataset (medium version)~\cite{FMA}. The original dataset contains \SI{25000} tracks of \SI{30}{\second}, in mp3 format from 16 unbalanced genres, at a sampling rate of \SI{44.1}{\kilo\hertz}. The tracks which original bit rate compression is lower than \SI{192}{\kilo\bit\per\second} are discarded. Each randomly selected track is converted to mono, cropped, and split into twenty-four  \SI{1103}{\milli \second}-long segments  for which the masking patterns are computed. Those 24 segments are put into the same randomly selected training or validation set to reach a total duration of respectively 80 and 10 hours.

\subsection{Training}

A set of models is trained with  \gls{mse} instead of \gls{nmr} in the loss formula  (\ref{eqn:loss}). For a message length of 256 bits, an initial model is trained for one epoch with  weight  ${\alpha=0.5}$. Then the weight is given its final value and the  training resumed for 29 more epochs. We choose five different final values for $\alpha$ (0.8, 0.9, 0.92, 0.95, and 0.98)  and train three models for each of them; if we  use directly the final weight values, without the initial epoch, the model would extract random messages. 
Those final values for $\alpha$ were chosen to get a set of models with various and decent acoustic properties and accuracy for the extraction so that all of them could be selected for the listening tests. 

A second initial model is trained with the \gls{nmr} for the three first epochs with successive weights $\alpha = 10^{-4}$, $10^{-2}$ and $10^{-1}$. Then three models are trained with each of the final values for $\alpha$ (0.1, 0.2, 0.3, 0.4, and 0.5) for 27 more epochs.  Those final weights values are chosen to get 15 \gls{nmr}  trained models   with \gls{ber} comparable to the 15 previously \gls{mse} trained models. 

Except the initial and final weights, and the distortion loss, all the models, with identical architecture, are trained in the same conditions (learning rate, batch size, Adam optimizer). The binary message is randomly drawn for each audio segment and at each epoch. The training sessions lasted at most 30 epochs, with early stopping when the validation loss did not improve for two consecutive epochs to prevent overfitting.  The model at the epoch with the best validation loss $\ell$ is selected. 
The balance between the accuracy of the message extraction and the transparency of the watermark depends on the final value of the weight~$\alpha$. For two models trained with the same loss function, a higher final value of $\alpha$ generally means a  better transparency (measured with \gls{mse} or \gls{nmr}) and a lower accuracy (measured with the \gls{bce}).

Out of those 15 models trained with \gls{mse}, we select the one with the best \gls{snr} on the test set.
Out of the 15 \gls{nmr} trained  models, we select the one with the best \gls{nmr} among those with lower \gls{ber}  than the  selected  MSE model. Those two models are the MSE256 and the NMR256 models of the Table \ref{tab:models}. 

The same scenario is reproduced with an input message of 512~bits, and we obtain the four models of the Table \ref{tab:models}. We can observe that for this longer messages the \gls{ber} is becoming very high and the \gls{snr} slightly worse and those models would not be accurate enough to be used for efficient audio watermarking.

\begin{table}[!ht] 
	\caption{The four selected trained models. The validation metrics used for the selection are in blue}
	\label{tab:models} 
	\sisetup{table-format=2.1, round-mode=places, round-precision=1}
	\centering
	\begin{tabular}{|c|S[table-format=1.5, round-precision=5]|S|S|} \hline 
		Model	 & {BER}  & {SNR (dB)} & {NMR (dB)}\\ \hline 
		MSE256  &  \color{blue}0.00030 &  \color{blue} 25.8 &  13.7 \\
		NMR256  & \color{blue}0.00025 & 22.4 & \color{blue} -18.0 \\
		MSE512  & \color{blue}0.111 & \color{blue} 25 & 13.7 \\
		NMR512  & \color{blue}0.097 & 18.6 & \color{blue} -13.3 \\ \hline 
	\end{tabular}
\end{table}

\subsection{Objective evaluation}

\begin{figure}[h]
	\centering	 		
	\includegraphics[width=\columnwidth]{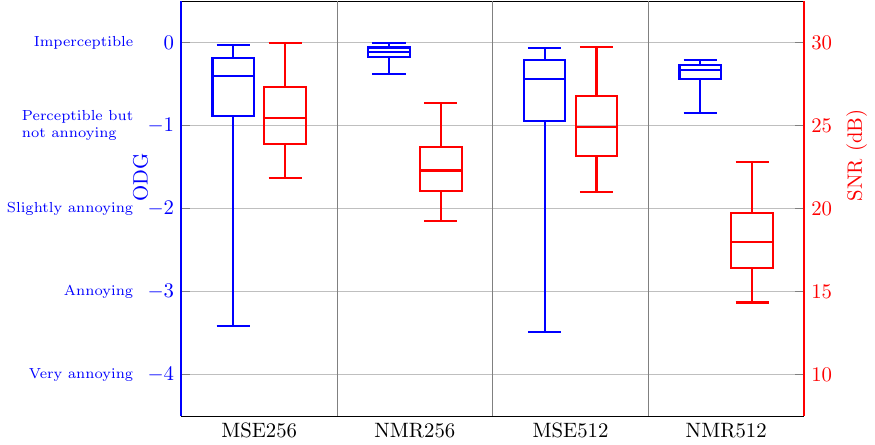}
	\caption{\gls{peaq} (\gls{odg} scale) in blue. The red boxes represent the \gls{snr}. The upper and lower whisker corresponds to the \nth{5}  and   the \nth{95} percentiles, the outliers are not shown  }
	\label{fig_objective_results}
\end{figure}

The objective evaluation is performed with the 100 tracks of the Music Genre Database~\cite{RWC}. For each track thirty 1103\,ms-long segments  are randomly selected. The \gls{peaq}~\cite{BS.1387}  is evaluated between the segments marked with each four selected models and the host segments. We used the MATLAB implementation by Kabal~\cite{kabal}. The results are presented on Fig.~\ref{fig_objective_results}.  

With the \gls{nmr}-models  more than 95\% of the 3000 segments get a \gls{peaq} rating between 0 (\emph{imperceptible}) and~$-1$ (\emph{perceptible but not annoying}).
For the \gls{mse}-models   around 25\% of the   segments   are rated between~$-1$ and~$-4$ (\emph{very annoying}). The segments marked by the two \gls{nmr}-models have a lower \gls{snr} in comparison to the ones marked by the \gls{mse} models, even though they have a better objective perceptual quality.  

\subsection{Subjective evaluation}

We organized a MUSHRA test~\cite{mushra} with  the platform  WEBMushra~\cite{webmushra} to evaluate the audio quality of the models of the  Table \ref{tab:models}. Although those test have been initially designed to assess audio coding they have been used  to evaluate the quality of watermarked audio~\cite{tachibana},~\cite{kondo} as well.

\begin{figure} 
	\centering 	
	\includegraphics[width=0.95\columnwidth]{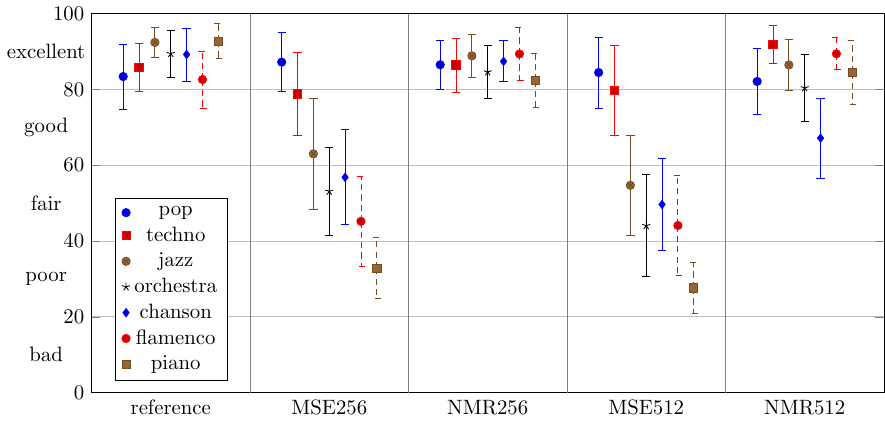}
	\caption{Mean and 95\% confidence intervals of the MUSHRA scores for the 7 clips and their versions marked with the four trained models}
	\label{fig:subjective_results} 
\end{figure}

We selected manually the 7 clips of \SI{10}{\second}  shown on Table~\ref{tab:clips} from the tracks used in the objective tests with the idea to present a variation of musical genre, number of instruments and spectral signature. For each clip the assessors had to evaluate blindly 7 versions: the original clip, two low-pass filtered anchors at \SI{3.5} and \SI{7}{\kilo\hertz} (recommended in MUSHRA tests), and the versions marked with the 4 trained models where the original clip is padded with zeros to the duration of 10 segments, the segments are embedded, concatenated and the result is cropped to the original \SI{10}{\second}-duration.  The 20 assessors were naive listeners,  between 15 and 50 years old, and had no previous experience in evaluation of audio quality. The results are shown on Fig.~\ref{fig:subjective_results}.  

\begin{table}[h]  
	\caption{The  10-second clips used for the MUSHRA test}
	\label{tab:clips}
	\centering
	\sisetup{table-format = 2.1, round-precision=0} 	
	\begin{tabular}{|l|l|l|S[table-alignment-mode = marker]|}
		\hline 
		Study ID & Filename & Title & {Start time (s)}  \\
		\hline
		pop & G001 & Wasting Time &  240 \\
		techno & G021 & Stf & 20 \\
		jazz & G030 & Kitchen  &58\\
		orchestra & G050 & Egmont & 210\\
		chanson & G087 & Je te Veux & 93 \\    	
		flamenco & G082 & Sevillanas & 70 \\
		piano & G059 & Rondo in D Major & 20 \\
		\hline 
	\end{tabular}   
\end{table}

For thirteen of the fourteen clips marked with the \gls{nmr}-models the average rate is \emph{excellent} (between 80\% and 100\%) whereas it is the case for only four of the clips marked with the \gls{mse}-models. For those, the two  original clips (\emph{pop} and \emph{techno}) are acoustically busy with multiple instruments playing at the same time. For the \emph{chanson} marked with the NMR512-model the quality was rated as \emph{good} showing that the watermark is not transparent, but the subjective evaluation is better than the \emph{fair} grade given for both MSE-models.  The worse ratings have been given to mono-instrumental clips (guitar and piano) marked with \gls{mse}-models.  

\section{conclusion}

In this study we propose a perceptual loss based on the \gls{nmr} which models the psychoacoustic masking effect of the human ear. We use this loss to train an autoencoder \gls{dnn} for music watermarking in the goal of minimizing the perceptual distortion between the host and the marked signal. We also train from scratch the same encoder with \gls{nmr} replaced by the commonly used \gls{mse} loss   
and we compare objective quality with \gls{peaq} and subjective quality with MUSHRA tests. Both tests show clear superiority  of the audio quality of the watermarked audio  embedded by models trained with our perceptual \gls{nmr} loss. 
Our solution is not dealing with robustness and synchronization, and this could be addressed by further studies where a \gls{dnn} watermarking system could be trained to resist to attacks and some mechanism found to tackle the synchronization issue. Our \gls{nmr} loss would be used in those future solutions to improve the transparency of the generated watermarks.

\bibliography{IEEEabrv,biblio}
	
\end{document}